\documentclass[journal=jpcck, manuscript=article]{achemso}
\setkeys{acs}{usetitle = true}
\usepackage{graphicx}
\usepackage[version=3]{mhchem} 
\usepackage{amsmath}
\usepackage{color}
\usepackage{epstopdf}
\usepackage{pdfpages}
\usepackage{color}

\author{Kim G.~L.~Pedersen}
\affiliation{Nano-Science Center}
\alsoaffiliation{Department of Chemistry}
\alsoaffiliation[Niels Bohr Institute]{Niels Bohr Institute, University of Copenhagen, Denmark}
\author{Anders Borges}
\affiliation{Nano-Science Center}
\alsoaffiliation{Department of Chemistry}
\author{Per Hedeg{\aa}rd}
\affiliation{Nano-Science Center}
\alsoaffiliation[Niels Bohr Institute]{Niels Bohr Institute, University of Copenhagen, Denmark}
\author{Gemma C. Solomon}
\affiliation{Nano-Science Center}
\alsoaffiliation{Department of Chemistry}
\author{Mikkel Strange}
\affiliation{Nano-Science Center}
\alsoaffiliation{Department of Chemistry}
\email{strange@chem.ku.dk}

\title{On the illusory connection between cross-conjugation and quantum interference}
\keywords{Quantum interference, conjugation, molecular electronics}

\begin{document}
\begin{abstract}
Quantum interference, be it destructive or constructive, has a substantial influence on the magnitude of molecular conductance and consequently there is significant interest in predicting these effects. It is commonly thought that cross-conjugated paths result in suppressed conductance due to destructive quantum interference. Using H\"{u}ckel theory and DFT calculations we investigate systems that break this cross-conjugation rule of thumb.  We predict and rationalize how a class of conjugated molecules containing closed loops can exhibit destructive interference despite being linearly conjugated and exhibit constructive interference despite being cross-conjugated. The arguments build on the graphical rules derived by \citet{mst_2010} and the hitherto neglected effects of closed loops in the molecular structure. Finally, we identify the 1,3 connected Azulene molecule as belonging to the closed-loop class and argue that this explains recent measurements of its electrical conductance.
\end{abstract}

\begin{figure}[tb]
\includegraphics[width=8.5cm]{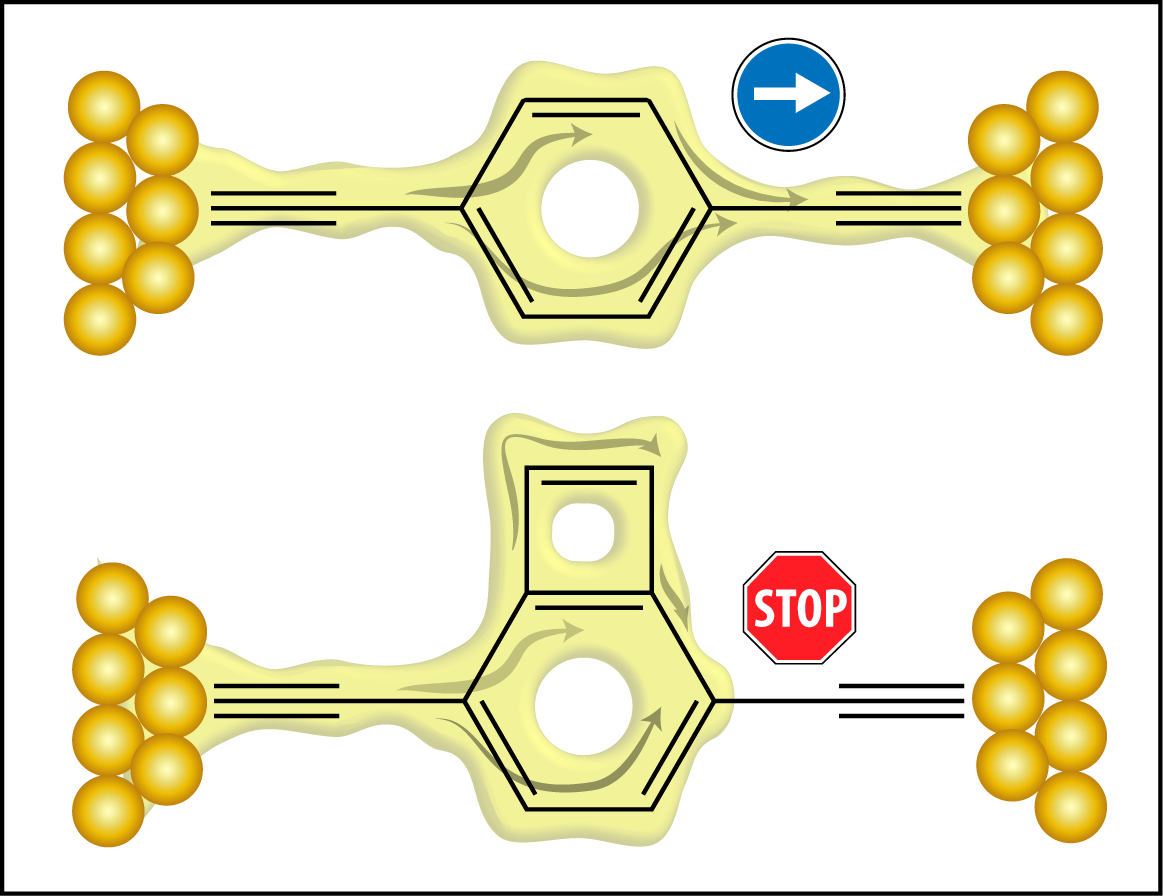}
\caption{\label{fig.toc} Table of contents graphic}
\end{figure}

\section{Introduction}
Structure-function relations are one of the cornerstones of chemistry, but as we venture into new types of measurements, for example measuring the current flowing through single molecules, we need to expand our understanding to encompass these types of systems. Initially, it seemed that this extension was not so complicated. Saturated systems had lower conductance than conjugated systems, and this could be understood in simple terms by the size of the gap between the highest occupied (HO) and lowest unoccupied (LU) molecular orbitals (MO). Recent interest in destructive quantum interference has refined this picture, with a focus on the nature of the path through the molecule~\cite{pcg_1997, gvm_2012, amv_2012, rcl_2013, atv_2013, xcc_2014}. A cross-conjugated or \textit{meta}-type substituted path will result in low conductance, on the order of a saturated system of the same length, while a linearly conjugated path will result in significantly higher conductance. We will refer to this idea as the ``cross-conjugation rule''~\cite{sar_2008a, vgh_2014}. Here we show that this is, however, not a complete picture, and both linearly- and cross-conjugated systems can exhibit destructive or constructive quantum interference.

It is desirable to have a readily applicable method to predict interference effects because of their great influence on conductance. Computational approaches range from a detailed description of the transport through a molecular junction, to calculations of the orbitals of the isolated molecule using either molecular orbitals (MOs) from mean-field calculations~\cite{yts_2008, th_2014, lm_2013, ssr_2014} or Dyson orbitals from higher level calculations~\cite{psp_2014}. However, in all cases, computational approaches necessarily require a computation and are therefore not ideal for developing a clear structure-function intuition.

An appealing alternative is to predict quantum interference directly from the topology of the molecule using the set of graphical rules introduced by Markussen, Stadler and Thygesen (MST)\cite{mst_2010, mst_2011}. This approach does not require any computation, beyond back-of-the-envelope type diagrams. Recent experiments on derivatives of an Azulene (Az) molecule suggest, however, that both the cross-conjugation rule and the graphical rules break down in this case~\cite{xcc_2014}. 

Here we show that this apparent breakdown is in fact an incomplete application of the graphical rules. We show that the systems where the cross-conjugation rule breaks down are limited to a class of molecular topologies where a cyclic moiety can be disconnected from the molecule when drawing a continuous tunneling path through the molecule. We call this the \emph{closed loop class}. We have performed both H\"{u}ckel theory and density functional theory (DFT) transport calculations for a few representative molecules to illustrate the illusory connection between quantum interference and conjugation. Finally, we argue that the experimental results for Az can now be rationalized, as some of the Az derivatives belong to this closed loop class of molecules. During the final stages of the preparation of this manuscript, we were made aware of a comment~\cite{s_2015} by one of the authors of the graphical rules, indicating that closed loops had been mistakenly omitted~\cite{xcc_2014} and that this resolved the issue of apparent breakdown of the rules for 1,3-substitued Az. We note in passing that there are other cases of apparent breakdown of the rules for Az~\cite{ssvc_2015} and that the complete set of graphical rules as outlined here resolves this issue. 

\section{Methods}
The H\"{u}ckel model includes nearest neighbor coupling terms ($t=-3$~eV) for the $\pi$-system only, with on-site terms $\alpha_i$ for site $i$. We set $\alpha_i$ for carbon atoms to $0$~eV. For the H\"{u}ckel model calculations, we do not include any binding group atoms, such as sulfur, but attach carbon atomic sites directly to wide-band leads. For the DFT calculations, we use the GPAW code employing a double $\zeta$ polarized basis set with a confinement energy of $0.01$~eV on Au and S and $0.1$~eV on C and H\cite{gpaw_2010,ctj_2012}. Exchange and correlation effects are included through the Perdew-Burke-Ernzerhof density functional~\cite{pbe_1996}.

The transmission is calculated as a function of energy using the standard Landauer-B\"{u}ttiker formula, ${T(E)=\text{Tr}[G^r(E)\Gamma_L(E) G^a(E) \Gamma_R(E)]}$, where $G^{r(a)}$ is the retarded (advanced) Green's function for the molecule coupled to leads. To include the effect of lead $\alpha$ on the molecule we use an embedding self-energy $\Sigma^r_\alpha$, which also defines the spectral broadening matrix, $\Gamma_\alpha=i(\Sigma^r_\alpha-\Sigma^a_\alpha)$. Throughout this paper we plot the transmission, noting that the conductance is proportional to the transmission at the Fermi energy ($E_F$) while the current is proportional to the integral of the transmission about $E_F$ over an energy window defined by the applied bias voltage. 

\section{Results and discussion}
\subsection{4 examples to illustrate the breakdown of the conjugation rule}
\begin{figure}[tb]
\includegraphics[width=8.5cm]{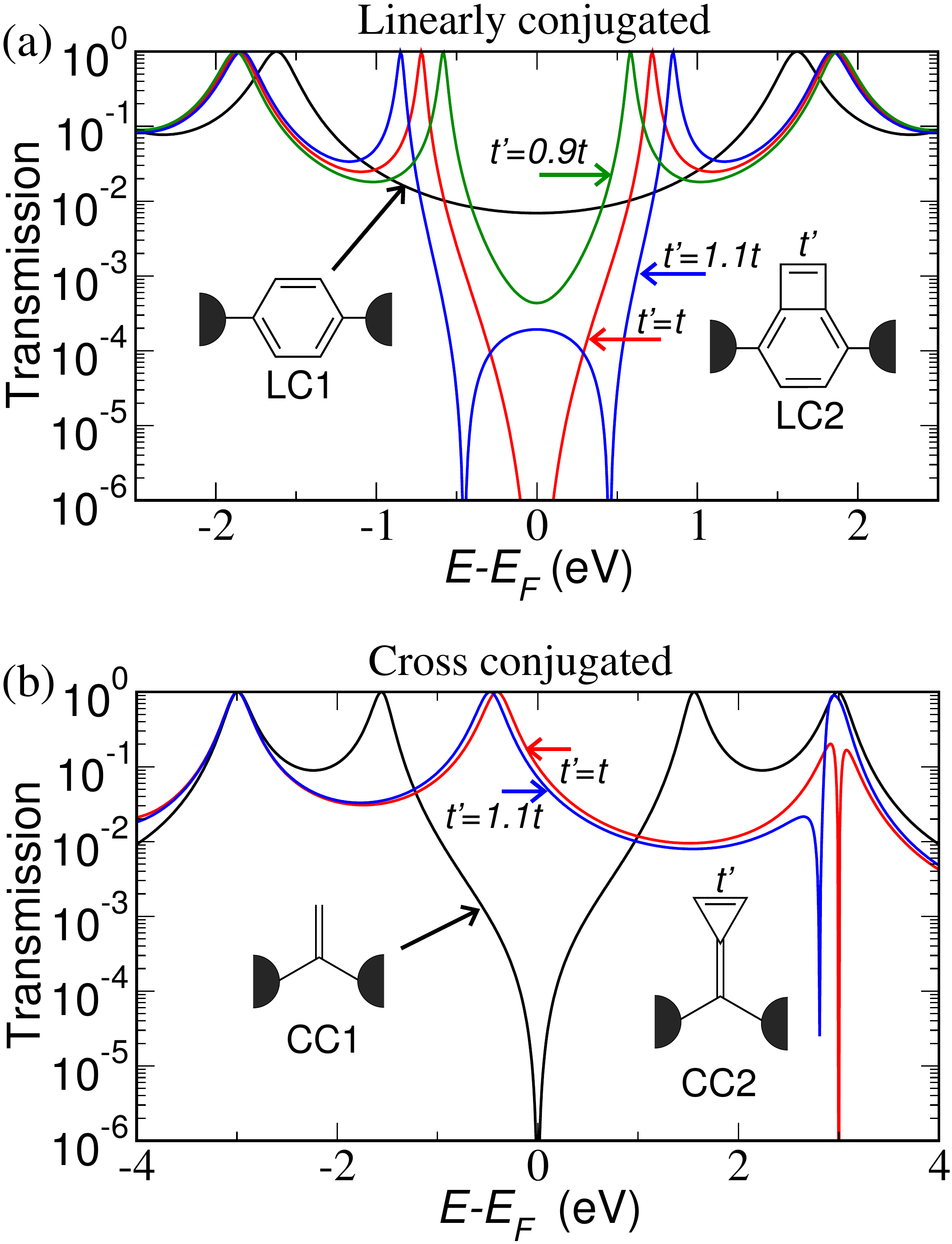}
\caption{\label{fig.fig1} Transmission for a H\"uckel model with hopping $t=-3$~eV between nearest neighbour atoms for pairs of (a) linearly-conjugated and (b) cross-conjugated molecules. The effect of bond length alternation is included by a modified hopping element $t'$ for the bonds indicated in the insets. 
}
\end{figure}

We have chosen four representative molecules to illustrate the connection, or rather lack thereof, between conjugation, molecular structure and quantum interference. The H\"uckel model structures for these systems are shown in Figure \ref{fig.fig1}: Two linearly conjugated structures, LC1 and LC2, and two cross-conjugated structures, CC1 and CC2. In each case, the first element of the pair is an archetypical example of either linear- or cross-conjugation and the second has an additional ring moiety on top of the basic structure. It is these additional rings that bring the systems into the closed-loop class and result in a dramatic change in the transport properties. 

The transmission for the four systems is shown in Figure~\ref{fig.fig1}, clearly showing how destructive quantum interference causes large dips in the transmission near $E_F$ for both types of conjugation (LC2 and CC1). Conversely, we also see no evidence of destructive interference near $E_F$ for the other members of each pair. The most simplistic implementation of the H\"uckel model treats all bonds as equal with no distinction for bond length alternation. If we relax this approximation for the LC2 and CC2 we see that while the transmission does change; the destructive interference feature(s) near $E_F$ remains. The effects of bond length alternation are more pronounced in LC2 than in CC2. Modeling a double bond using $t'=1.1t$ has a substantial effect on the transmission (blue line), which is still small compared with LC1 but no longer goes to zero at the Fermi level. Instead there is now two split interference features at $\sim\pm0.5$~eV. When using $t'=0.9t$ to model a single bond, the transmission has no nodes (green line), but is still highly suppressed at the Fermi level compared with LC1. We speculate that this highly suppressed transmission is related to complex solutions of the polynomial determining 
the node positions and the broader definition of destructive interference introduced by Reuter and Hansen\cite{rh_2014}.

\begin{figure}[tb]
\includegraphics[width=8.5cm]{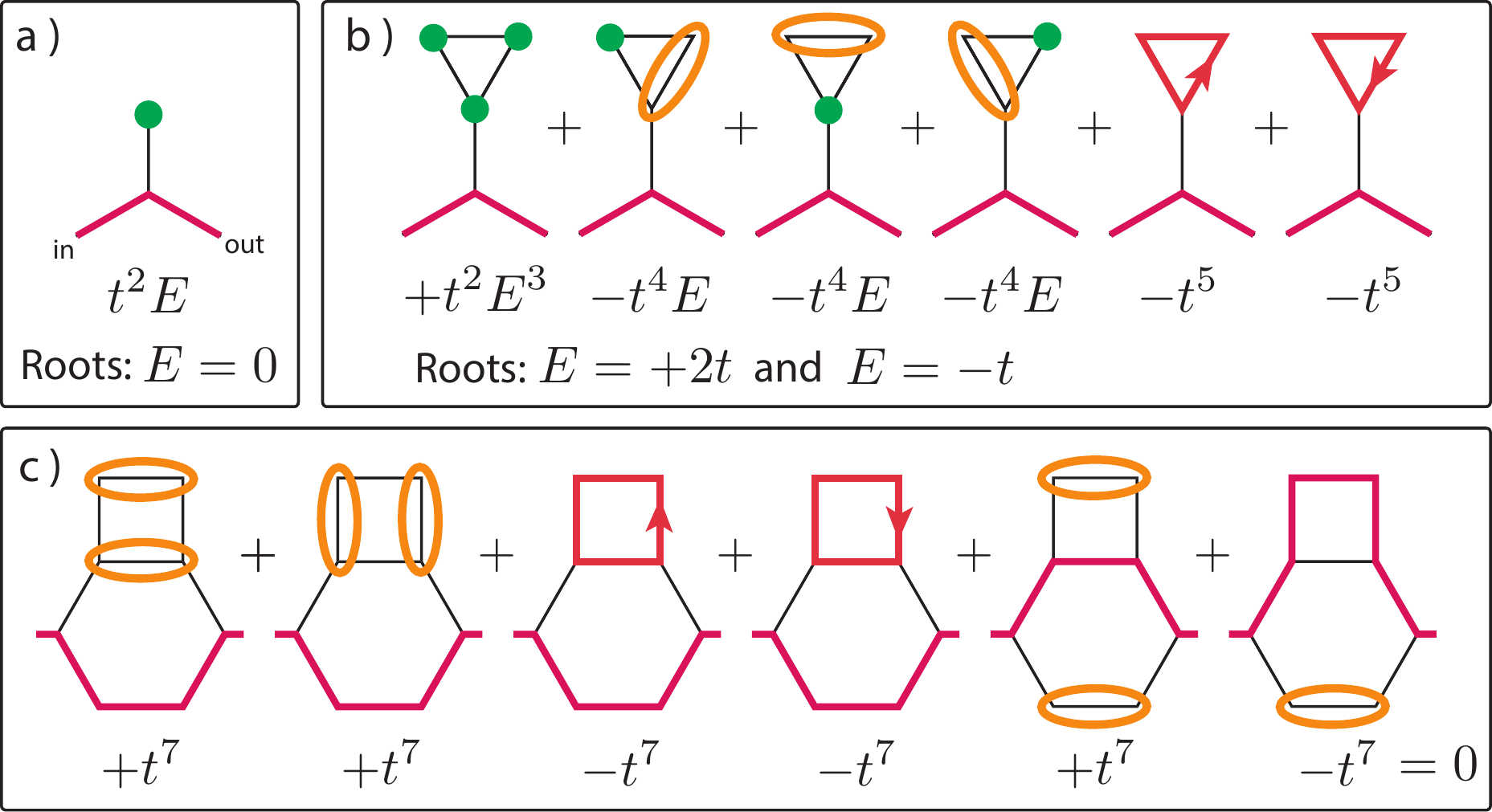}
\caption{\label{fig.fig2} Markussen diagrams for molecule (a) CC1 and (b) CC2. The contribution from each 
diagram to the characteristic polynomial, which has roots at the transmission node positions, 
is shown directly below it. The two rightmost diagrams in (b) are of the closed loop form and responsible for 
switching off the interference for the cross conjugated molecule CC2.  
(c) All diagrams for molecule LC2 which does not contain an on-site loop is seen 
to cancel out. This results in the linearly-conjugated molecule LC2 to have an interference feature at $E=E_F$.}
\end{figure}

\subsection{Understanding the breakdown using graphical rules}
The cross-conjugation rule clearly breaks down for these examples where the paths between the electrodes remain unchanged for the two pairs but the interference effects change significantly. Next, we rationalize these finding using the graphical rules. 
The graphical rules predict transmission nodes and are valid for a H\"{u}ckel  description where each lead is only connected to a single atomic site of the molecule. In this case, the transmission can be written as $T(E)=\Gamma(E)^{2}|G^r_{1N}(E)|^2$, where $1$ and $N$ denote the atomic sites connected to the leads. The energies of the transmission nodes can then be obtained from the equation $G_{1N}(E)=0$. A derivation of the graphical rules can be found in Reference~\citenum{mst_2010}; 
but we have, for completeness, included a re-derivation including some relevant details in the Supporting Information. All possible graphical diagrams are drawn from the rules that have been previously outlined \cite{mst_2011}, with some additional details required to turn these diagrams into a full sum over terms. For a molecule connected to leads through sites $1$ and $N$, described by a H\"uckel Hamiltonian with nearest-neighbor hopping elements $t$ and on-site terms $\alpha_i$, the rules can be outlined as:
\begin{enumerate}
\setcounter{enumi}{0}
  \item Draw a path (traversing $L$ bonds) that connects the two sites $1$ and $N$ through hopping elements in the Hamiltonian. The path contributes a factor of $t^L$.
  \item For each path, draw all possible combinations for the remaining sits using:
  \begin{enumerate}
    \item[(A)] On-site loops marked by a green circle and corresponding to a factor of $E-\alpha_i$.
    \item[(B)] Pair loops joining two neighboring sites contributing a factor of $t^2$ and drawn as an ellipse. 
    \item[(C)] Closed loops (length $n>2$) with a given orientation (clockwise or anti-clockwise) contributing a factor of $t^n$.
    \end{enumerate}
\item The sign of each term is $(-1)^p$, where $p$ is the number of pair loops and closed loops, i.e. loops encircling more than one site.
\end{enumerate}
We note that the rules above use a different choice than the MST rules for determining the sign of a diagram.
For more details about the different choices for calculating the signs, we refer to the Supporting Information.


Several applications of the graphical rules can be found in Reference~\citenum{mst_2010} and \citenum{mst_2011}, but the molecules we discuss in this work give rise to diagrams which have not been treated explicitly previously.  

We show in Figure \ref{fig.fig2} the relevant diagrams for CC1, CC2 and LC2. The graphical analysis of LC1, which shows constructive interference, is straightforward and can be found in Reference \citenum{mst_2010} and in the Supporting Information. According to the rules above we draw diagrams using green dots to represent on-site loops (factor of $E$) red lines to represent a path (factor of $t$ for each bond traversed) and pair loops are represented by an orange ellipse (factor of $t^2$). Below each diagram, we show the contributions to the polynomial and the corresponding root(s). 

Because there is only one diagram for CC1, the polynomial becomes particularly simple, $(-1)^{0} t^2 E$, with a root $E=0$~eV in agreement with the transmission shown in Figure \ref{fig.fig1}b. For CC2, roots are found at $E=2t=-6$~eV and $E=-t=3$~eV; the latter node is visible in the transmission in Figure~\ref{fig.fig1}b, 
while the former is outside the plotted range. While most of the diagrams for CC2 contain at least one green dot, the two rightmost diagrams do not. Instead they contain what we call a closed loop path, one going clockwise and another going anti-clockwise. 

These diagrams have not been considered explicitly before, but they will show up whenever a path going through the molecule leaves at least one cyclic fragment behind. Indeed, if we leave out the closed-loop diagrams for CC2, the remaining ones all have a common factor of $E$ (green dot), which implies a node at $E=0$~eV in accordance with the cross-conjugation rule. 
But including the closed loop path diagrams, where each diagram contributes a constant value ($t^5$), means that $E=0$~eV is no longer a root. 
This observation for CC2 ican be generalized such that using a restricted set of diagrams, where the closed loop path diagrams are excluded, corresponds exactly to the cross-conjugation rule for predicting quantum interference, 
see Supporting Information for more details. 

To rationalize the destructive interference feature in the linearly-conjugated molecule LC2, we show all diagrams that do \emph{not} contain on-site loops in Figure~\ref{fig.fig2}c. Note that the contributions to the polynomial from these diagrams exactly cancel. This means that $E=0$~eV is a root, since $E$ is a common factor of all remaining diagrams. As seen in Figure~\ref{fig.fig1}a, this is in agreement with the transmission for LC2 (without bond length alternation). We note that the cancellation relies on the inclusion of the closed loop path diagrams (3rd and 4th diagram from left) for LC2. Furthermore, the rather large sensitivity of the transmission to bond length alternations can easily be understood from the diagrammatic structure. The transmission node at $E=0$~eV depends on the cancellation of diagrams that contain a number of hopping elements originating from different bonds in the molecule. When hopping elements vary from 
bond to bond, the cancellation of the diagrams are in general not complete and $E=0$~eV is not a root of the polynomial. 

\subsection{Correspondence with DFT calculations}
The graphical rules give a particularly elegant way of understanding the relation between molecular structure and quantum interference and reveal when the cross-conjugation rule works and when it does not. However, the graphical rules are based on H\"{u}ckel theory and the question arises as to what extent the predictions carry over to more sophisticated theoretical treatments and ultimately to experiment. 
\begin{figure}[tb]
\includegraphics[width=8.5cm]{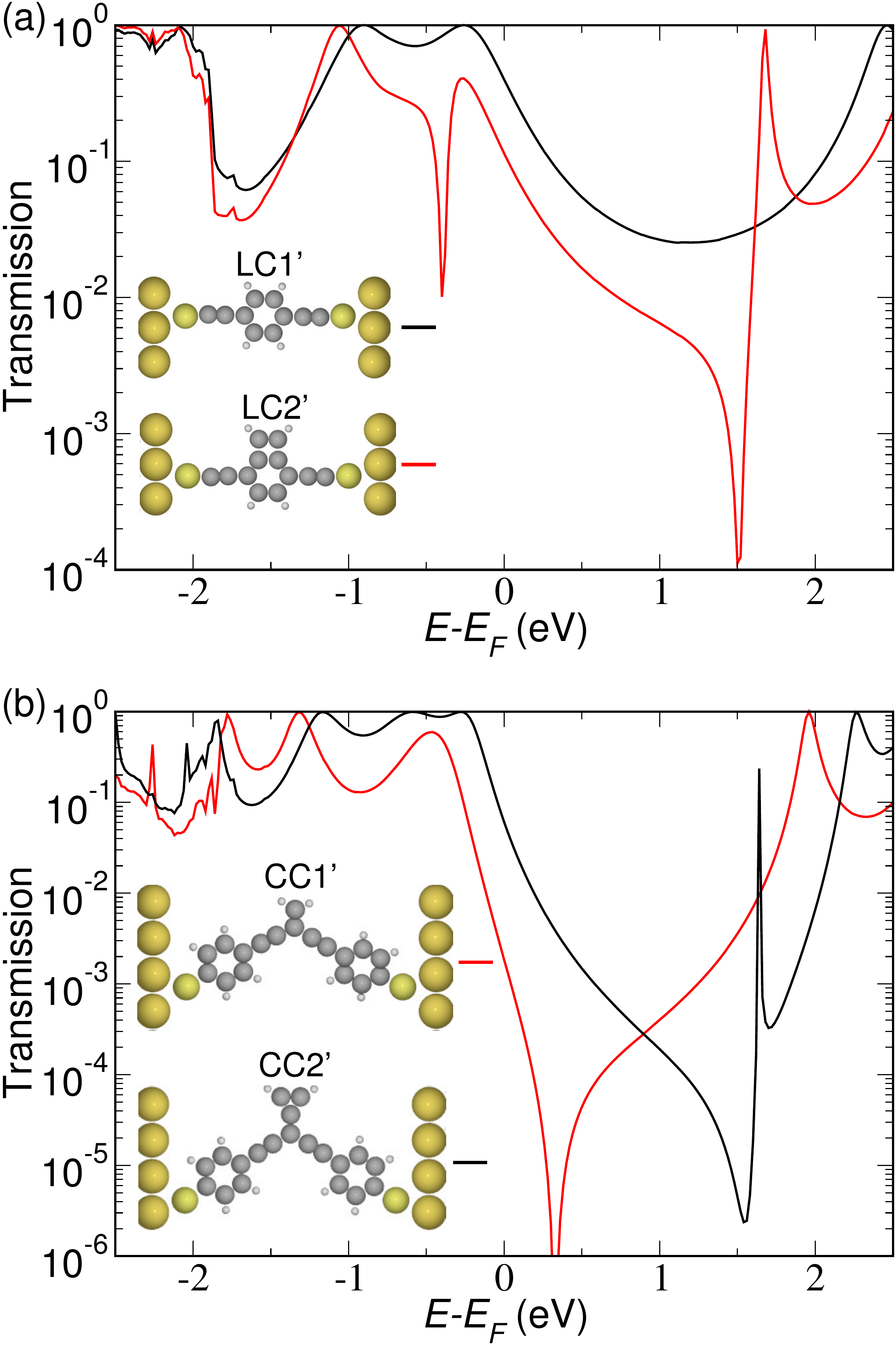}
\caption{\label{fig.fig3} (a) Transmission calculated with DFT for the molecules LC1' and LC2' which are linked to gold using 
ethynyl-thiol groups. (b) Transmission for the molecules calculated with DFT for CC1' and CC2' which uses ethynyl-phenyl-
thiol linker groups. The inset illustrates the geometry used in the calculations.}
\end{figure}

To address this question, we have performed DFT transport calculations for four molecules LC1', LC2', CC1' and CC2'. Each has the same central core as LC1, LC2, CC1 and CC2, respectively, but include spacers and thiol anchoring groups. 
We use ethynyl spacers for LC1' and LC2', while we use para connected phenyl spacers for CC1' and CC2' in order to suppress transport through the sigma-system associated with the triple bond. The optimized structure for LC2' has clear bond-length alternation, similar to the $t'>t$ case we examined in the model system calculation. 

We show in Figure \ref{fig.fig3} the transmission for the four systems calculated with DFT. The inset shows the space-filling structure used to model the molecular junction. In all cases, the interference features observed in the model system calculations are preserved in the DFT calculations; however the presence of additional orbitals from the binding groups change some details. The HOMO in these systems is now a thiol-dominated orbital which can shift the positions of the interference features (e.g. CC1') or even result in an interference feature no longer being in the HOMO-LUMO gap (e.g. LC2'). Additionally, in cases where the broadening of resonances differs between the H\"uckel calculations and DFT calculations (e.g. the LUMO in CC2') the range of energies with suppressed transmission can change accordingly. The thiol anchoring group has a significant influence on the frontier orbitals and consequently a significant impact on the interference effects. A binding group with weaker coupling to the molecular states, for example amine or methyl-thiol, may improve the quantitative agreement between the predictions from the graphical rules and the DFT results. The fact that the general conclusions remain, however, implies that the predictions from the graphical rules are rather robust at the mean-field level of theory.
 
 \subsection{Correspondence with experiment}
While DFT calculations are certainly more sophisticated than H\"{u}ckel theory calculations, they too have their limitations. Ultimately, we would like to compare these predictions with experiments and conveniently there exists a precedent for the closed-loop class in the literature. A recent experiment on derivatives of Az -- where the position of anchoring groups was varied systematically -- showed a conductance pattern that is not compatible with simple interference rules~\cite{xcc_2014}, including the cross-conjugation rule of thumb. In particular, the derivative denoted 1,3-Az gave the highest conductance of all derivatives although graphical rules neglecting loop diagrams suggested destructive QI~\cite{xcc_2014}. Based on the diagrammatic analysis described above, the 1,3-Az molecule can now be identified as a closed-loop type where additional diagrams are important for predicting (the absence of) destructive interference.
\begin{figure}[tb]
\includegraphics[width=8.5cm]{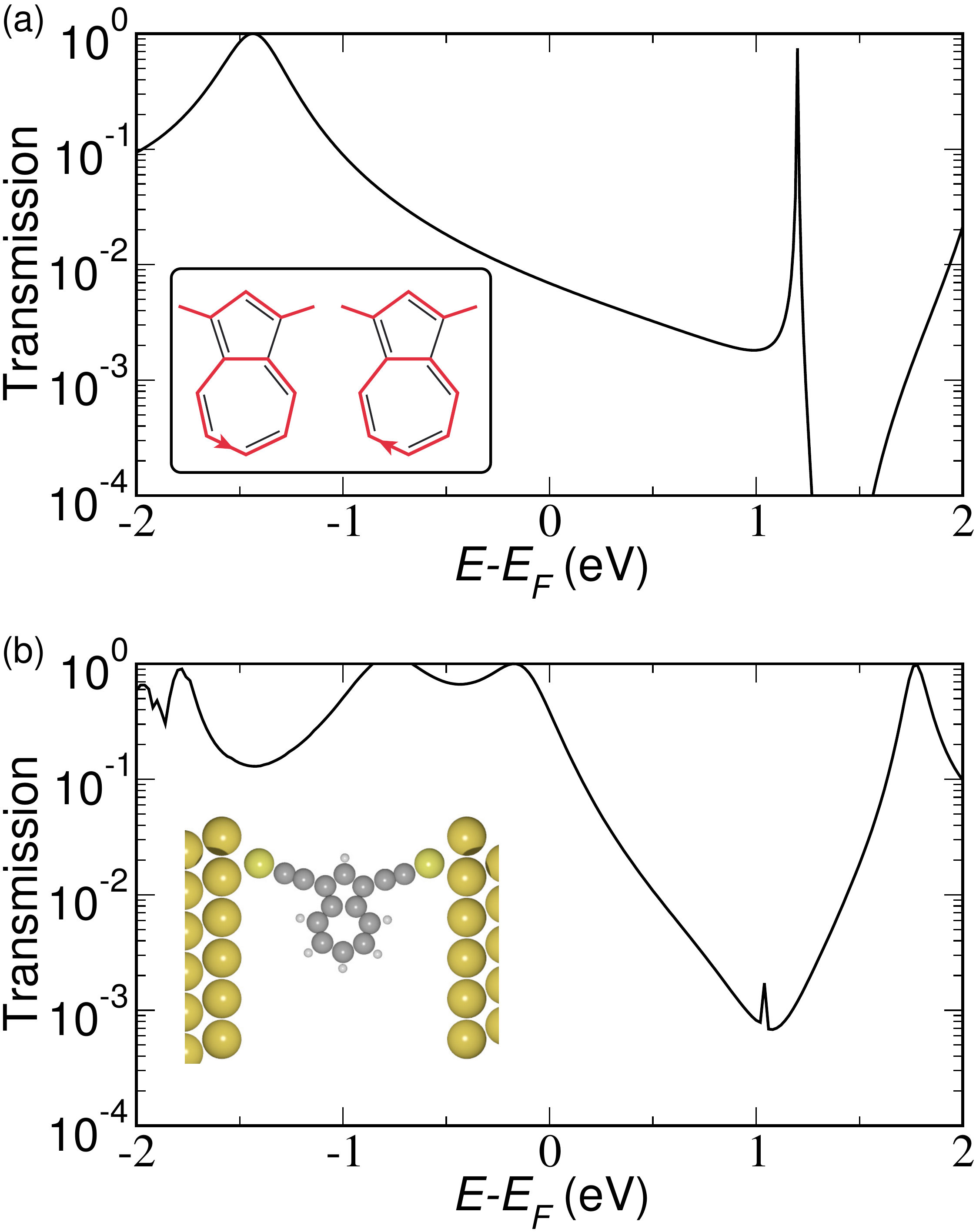}
\caption{\label{fig.fig4} (a) H\"{u}ckel transmission and (b) DFT transmission for Az-1,3. The inset in (a) shows the diagrams which do not contain an onsite-loop and shifts the interference feature away from the Fermi level. The same trend is observed in the DFT calculations in (b). Here the inset shows the geometry.}
\end{figure}

We show in Figure \ref{fig.fig4}a the transmission for 1,3-Az calculated with the H\"{u}ckel model. The inset shows a chemical drawing of the molecule as well as the important closed loop path diagrams. They come in pairs, one going clock-wise the other going anti-clockwise, as discussed above. Since these two closed loop diagrams are the only diagrams that do not contain on-site loops, we conclude that they are the reason why there are no transmission nodes within the HOMO-LUMO gap. Indeed leaving out the closed loop diagrams, \emph{i.e.} using the restricted set of graphical rules, one would predict a transmission node at $E=0$~eV~\cite{xcc_2014}.

We show in Figure \ref{fig.fig4}b the transmission from DFT and in the inset the geometry used to model the junction. The transmission shows qualitatively the same features within the HOMO-LUMO gap as the H\"{u}ckel model, in particular there are no transmission nodes. 
We note that the calculations for 1,3-Az in Reference \citenum{xcc_2014} use a more sophisticated description of exchange 
and correlations effects, which results in two transmission nodes just below the LUMO resonance. This does 
not change the qualitative picture for most energies within HOMO-LUMO gap. It does, however, serve to show that 
quantitative quantum interference predictions may be sensitive to the level at which exchange and correlation are described~\cite{sbr_2011, psp_2014}. Also, we note that the graphical rules are usually applied to predict interference features in the middle of the gap, but there are `exceptions' to this in the sense that some molecules may have split interferences already at the H\"{u}ckel level of treatment~\cite{yts_2008}.

\section{Conclusions}
In summary, we find that the relation between conjugation, molecular structure and quantum interference can be rationalized in terms of a diagrammatic method based on H\"{u}ckel theory~\cite{mst_2010,mst_2011}. This method reveals that when molecules contain certain closed-loop moieties, cross-conjugation need not yield destructive interference. This is explained by the appearance of diagrams with a closed loop structure that have not been considered previously. Conversely, if a molecule is not within this closed loop class, then cross-conjugation is very likely to give destructive interference. In general, even-alternant closed-shell molecules without closed loops are the least likely to present surprising transport properties and care should be taken when molecules have closed loops, are open shell (for example diradicals) or are non-alternant. 

A closed-loop molecule can be identified as having disconnected ring fragments when drawing a single path that connects the leads. We show that a cross-conjugation rule for hydrocarbons is fully equivalent to the graphical method when using a subset of diagrams where the closed loop path diagrams are omitted. 

From a diagrammatic point of view, the destructive quantum interference for linearly-conjugated molecules differs from that of cross-conjugated molecules. For a linearly-conjugated molecule the destructive interference relies on a cancellation of a number of diagrams. Since these diagrams contain hopping elements originating from different bonds in the molecule, the cancellation is sensitive to bond length alternations.
For the linearly-conjugated molecule LC2, the closed loop diagrams are essential for the cancellation leading to destructive 
quantum interference. However, destructive interference based on cancellation of diagrams is not restricted to the closed loop class and could occur in other linearly conjugated molecules. 

\begin{acknowledgement}
This work was supported by The European Union Seventh Framework Programme (FP7/2007-2013) under the 
ERC grant agreement no. 258806 and the Danish Council for Independent Research | Natural Sciences.
\end{acknowledgement}



\providecommand{\latin}[1]{#1}
\providecommand*\mcitethebibliography{\thebibliography}
\csname @ifundefined\endcsname{endmcitethebibliography}
  {\let\endmcitethebibliography\endthebibliography}{}



\section*{Supplementary material}

# Supporting information for:

# On the illusory connection between cross-conjugation and quantum interference

Kim G. L. Pedersen,[†,‡,¶] Anders Borges,[†,‡] Per Hedegård,[†,¶] Gemma C. Solomon,[†,‡] and Mikkel Strange[*,†,‡]

†Nano-Science Center

‡Department of Chemistry

¶Niels Bohr Institute, University of Copenhagen, Denmark

E-mail: strange@chem.ku.dk

## Derivation of the graphical rules

We assume a molecule described by a simple Hückel model with $N$ atomic sites. The molecule is coupled symmetrically to a set of wide-band electrodes only through the atomic sites labelled 1 and $N$. For this model, the Landauer transmission $T(E)$, through the molecular junction takes the form

$$T(E) = \gamma(E)^2 |G_{1N}(E)|^2. \quad (1)$$

Here, $G_{1N}$ is the $1, N$-th element of the molecular Green's function and $\gamma(E) = 2\Gamma_L(E) = 2\Gamma_R(E)$ describes the coupling to the leads through the self energies $\Sigma_L$ and $\Sigma_R$ for which $\Gamma_{L/R} = -2Im(\Sigma_{L/R})$. From Cramer's rule[S1] the $1N$ matrix element of the Green's function becomes

$$G_{1N}(E) = \frac{(-1)^{N+1} \det_{N1}(E - H_{\text{mol}})}{\det(E - H_{\text{mol}} - \Sigma_{\text{L}} - \Sigma_{\text{R}})}, \quad (2)$$

where $\det_{N1}(M)$ is the sub-determinant where the N'th row and 1'st column are removed from $M$. We note that the self energies $\Sigma_L$ and $\Sigma_R$ do not appear in the numerator as they drop out when taking the sub-determinant.

Complete destructive interference occurs when the transmission $T(E)$ vanishes. This happens when the numerator of Eq. (2) is zero:

$$\det_{N1}(E - H_{\text{mol}}) = 0 \tag{3}$$

We assume that the denominator of Eq. (2) is non-zero at any such energy. In the weak coupling limit the zeros of the denominator lie close to the eigenenergies of the Hamiltonian describing the isolated molecule.

We use Leibniz's rule to evaluate the sub-determinant. Leibniz' rule allows us to express the determinant of the matrix $A = E - H_{mol}$ in terms of the matrix elements $A_{ij}$ and permutations $\sigma$ of the site indices,

$$\det(A) = \sum_\sigma \text{sgn}(\sigma) \prod_i A_{i\sigma_i}. \tag{4}$$

In order to simplify the derivation we calculate the $N1$ sub-determinant as the full determinant of a matrix $A^{N1}$, with the matrix elements in the first column and $N$th row are replaced by zeros except the $N1$ element which is set to 1.

The expression for Eq. (3) can be evaluated for each permutation $\sigma$, which is written as the product of disjoint permutation cycles (including permutation fixed points).

All non-zero contributions to the sub-determinant contain a permutation $\sigma = (1\ i_1\ i_2\ \ldots\ i_{L-1}\ N)$, which correspond to a path of length $L$ connecting site 1 to $N$. The corresponding matrix element is given as, $(-t_{1i_1})(-t_{i_1 i_2})\cdots(-t_{i_{L-1}N}) \sim (-t)^{n-1}$. Orbital sites not included in this path are covered by closed permutation cycles: one-cycles (permutation fixed points), referred to as on-site loops in the main text, contributing a factor of $E$, and $n$-cycles $l_1 \mapsto l_2, l_2 \mapsto l_3, \ldots, l_n$ contributing a factor of $(-t_{l_1 l_2})(-t_{l_2 l_3})\cdots(-t_{l_n l_1}) \sim (-t)^n$.

According to a standard result in linear-algebra, the *parity* of the permutation $\sigma$ can be calculated in terms of the number of disjoint cycles $N_\sigma$ and the total number of orbital sites $N$, such that $\text{sgn}(\sigma) = (-1)^N (-1)^{N_\sigma}$.

We now possess all the necessary ingredients for evaluating the transmission nodes of Eq. (3). Each term of the sub-determinant is on the form $(-t)^a E^b$ with $a+b = N-1$. Energies at which destructive interference occurs is found by solving the resulting polynomial equation, $\det_{N1}(E-H) = 0$. Mathematically the derivation is straightforward, but tends to disguise the usefulness of the method. To illustrate the procedure we show a simple example.

Benzene (here ignoring linker groups) has a Hückel Hamiltonian (with $t \approx -2.5$ eV) given by

$$H_{\text{benz}} = \begin{bmatrix} 0 & t & 0 & 0 & 0 & t \\ t & 0 & t & 0 & 0 & 0 \\ 0 & t & 0 & t & 0 & 0 \\ 0 & 0 & t & 0 & t & 0 \\ 0 & 0 & 0 & t & 0 & t \\ t & 0 & 0 & 0 & t & 0 \end{bmatrix}, \text{ for } \quad \tag{5}$$

We assume that this molecule is coupled to electrodes in the meta position (through orbitals 1 and 3), and the relevant sub-determinant then takes the form

$$\det_{31}(E - H_{\text{benz}}) = \det \begin{bmatrix} 0 & -t & 0 & 0 & 0 & -t \\ 0 & E & -t & 0 & 0 & 0 \\ 1 & 0 & 0 & 0 & 0 & 0 \\ 0 & 0 & -t & E & -t & 0 \\ 0 & 0 & 0 & -t & E & -t \\ 0 & 0 & 0 & 0 & -t & E \end{bmatrix}.$$

To evaluate this determinant, we shall consider all permutations of the indices $1, \ldots, 6$. Let us take a example permutation,

$$\sigma = (1 \mapsto 3, 2 \mapsto 1, 3 \mapsto 2, 4 \mapsto 5, 5 \mapsto 4, 6 \mapsto 6)$$
$$= (123)(45)(6),$$

where the last line has been rewritten as a product of three disjoint cycles. The calculation is presented graphically in Figure S1. We find that the sign of the permutation is $\text{sgn}(\sigma) = (-1)^6(-1)^3 = -1$ and the factor evaluates to $(-t)^4 E$. Summing all possible permutations, we find, $\det_{31}(E - H_{\text{benz}}) = t^2 E^3 - t^4 E$, which indeed has a node at $E = 0$.

The example illustrates how terms of the sub-determinant $\det_{N1}(E - H)$ can be represented as molecular diagrams. This makes it possible to evaluate the sub-determinant graphically by constructing molecular diagrams according to the recipe:

1. Draw a path (length $L$) that connects the two sites 1 and $N$ through hopping elements in the Hamiltonian. The path contributes a factor of $(-t)^L$.

2. For each path, draw all possible combinations of the remaining orbitals using:

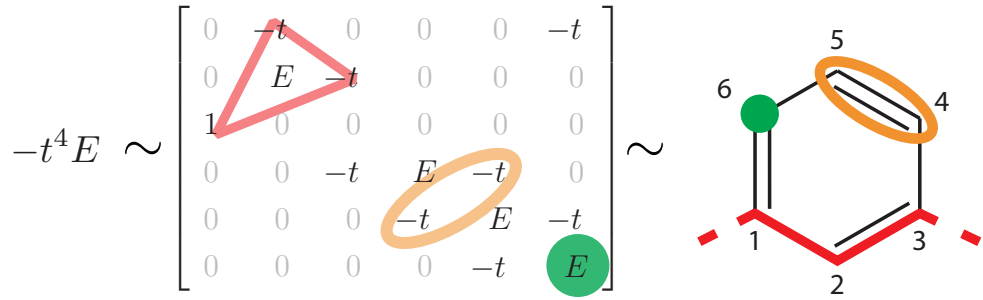

Figure S1: An example showing the contribution to the determinant from the permutation $\sigma = (123)(45)(6)$ in meta-coupled benzene. Note the correspondence between the subdeterminant and the molecular diagram.

(A) Fixed points (lone sites) marked by a green circle and corresponding to a factor of $E - \alpha_i$, where $\alpha_i$ is the on-site energy.

(B) Neighbor pairs joined by a 2-cycle contributing a factor of $(-t)^2$ and drawn as an ellipse.

(C) Closed loops (length $n > 2$) with a given orientation (clockwise or counterclockwise) contributing a factor of $(-t)^n$.

3. Each term comes with an additional sign $\text{sgn}(\sigma) = (-1)^N(-1)^{N_c+1}$, where $N_c$ is the total number of cycles, i.e. on-site loops, pair loops and closed loops.

Note that in 3. above we use that $N_\sigma = N_c + 1$, since the path from site 1 to $N$ is always part of a closed loop cycle that we do not draw and hence has to be added to $N_c$.

The rules above and especially the determination of the signs are closely related to the rules in Ref.[S3] We would like to note, however, that the factor of 2 coming from rule C above was never a part of the MST rules, although it clearly a part of the mathematics.

It is somewhat cumbersome to get the correct sign of each term with this recipe. The rules in the main text use a different scheme where all the signs are combined into one single factor. To show that the two ways of calculating the signs are equivalent, we first note that for a diagram containing a factor of $E^a(-t)^b$ has $a$ on-site loops per definition. If we write the number of cycles in a diagram as $N_c = a + p$, where $p$ is the number of loops encircling

more than one site, then a diagram (including all sign factors) can written as

$$\begin{align}
\text{sgn}(\sigma)E^a(-t)^b &= (-1)^N(-1)^{a+p+1}E^a(-t)^b \\
&= (-1)^{N+a+p+1}E^a(-t)^{N-1-a} \\
&= (-1)^{2N+p}E^a t^b \\
&= (-1)^p E^a t^b, \tag{6}
\end{align}$$

where we in the second line used that $b = N - 1 - a$.

If every term in the sub-determinant contains at least one fixed point, the transmission $T(E) \propto \det_{N1}(E - H) \propto E$ vanishes at a Fermi energy $E = 0$ due to destructive quantum interference. These graphical rules for predicting destructive interference at $E = 0$ in Hückel models have been investigated by Markussen et al.[S2,S3]. Here we make the distinction between the "*original* graphical rules" (rules 1,2A and 2B) published in Markussen et al.[S2] and the "graphical rules" (all rules 1,2,3) of Markussen et al.[S3].

The prediction of destructive quantum interference for meta-coupled benzene also follows from the cross-conjugation rule, where (when adding linker groups in the meta-position) the two meta-connected orbitals are *only* connected through cross-conjugated paths.

It is in fact generally true that the original graphical rules (ignoring the closed-loop-rule) produce predictions identical to the cross-conjugation rule. The closed-loop rule (2C) and the parity rule (3) make it possible to systematically break the cross-conjugation rule (at least at the Hückel level of theory).

# The equivalence of the cross-conjugation rule and the original graphical rules

Next we prove that the original graphical rules (rules 1, 2A and 2B) and the cross-conjugation rule, both give the same predictions regarding destructive interference. Note that we consider only molecules with a simple molecular graph, i.e. hydrocarbon molecules with an even number of orbitals in the π-system.

We will specifically prove that the original graphical rules predict a transmission node at $E = 0$ (destructive quantum interference), if and only if all paths through the molecule are cross-conjugated.

We break the proof down into two statements. 1) There exist graphical diagrams with no fixed points (i.e. no transmission node), if and only if there is a conjugated path through the molecule. 2) For each cross-conjugated path, one can only construct graphical diagrams

containing at least one fixed point (i.e. a transmission node), unless there exists an alternative linearly conjugated path through the molecule.

A path through the $\pi$-system of the molecule can traverse a double bond either longways or sideways. This is illustrated in Figure S2.

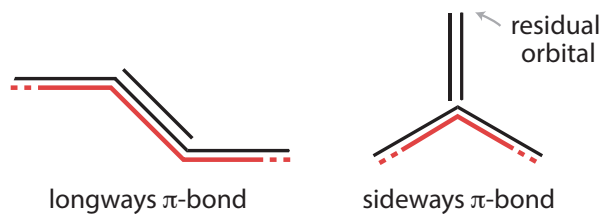

Figure S2: A path traveling longways (left) or sideways (right) through a double bond. Sideways traversal implies that the path is cross-conjugated. In the original graphical rules the residual orbital (not traversed by the path) must be covered by a (green) fixed point. Hence both methods predict destructive quantum interference for paths which traverse at least one $\pi$-bond sideways.

A path traversing all double bonds longways is conjugated. Since such a path traces out only double bonded pairs, the untraversed atomic sites can be covered by graphical rule pairs (rule 2b) by simply overlaying the remaining double bond structure. This proves the first statement.

Let us then turn to the second statement. A cross-conjugated path must traverse at least one $\pi$-bond sideways. Any cross-conjugated path which traverses an odd number of double bonds in a sideways fashion, leaves an odd number of residual orbitals. Because the odd number of remaining sites cannot be covered by pairs only, the second statement holds.

Cross-conjugated paths traversing an even number of double bonds in a sideways fashion, visits an even number of sites. Assume that the remaining sites can be completely covered with graphical pairs only (see e.g. Figure S3). We can then construct an alternative molecular graph, i.e. a resonance. This can be done by drawing double bonds between graphical pairs which allow us to turn the cross-conjugated path into a linearly conjugated one (because the path visits an even number of sites). This proves the second statement and therefore proves that the cross-conjugation rule can be used interchangeably with the original graphical rules (1, 2A and 2B).

## The failure of the cross-conjugation rule

Because the cross-conjugation rule is only identical to a subset of the graphical rules, we can break the cross-conjugation rule by the applying the additional graphical rules (2C and 3).

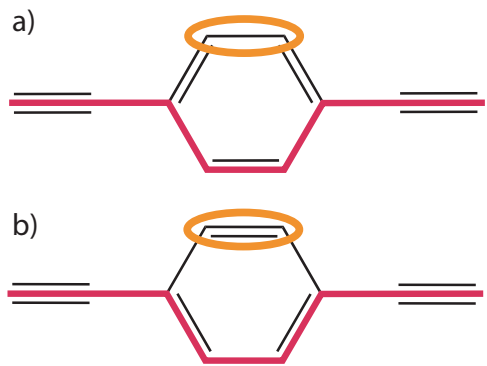

Figure S3: An alternative pair covering for a path traversing an even number of $\pi$-bonds in the sideways fashion. **a)** The drawn path is clearly cross-conjugated but the remaining atoms can be paired. **b)** When we redraw resonance structure a linearly conjugated path is evident.

Rule 2C involves closed loops. The presence of a closed loop outside the main path can change the position of a predicted transmission node away from $E = 0$. For azulene the node is pushed out of the HOMO-LUMO gap. Rule 3 involves the sign of sub-determinant terms, which could result in the cancellation of different graphical diagrams. Such cancellations usually change the prediction about interference for non-alternant molecules (like Azulene, see Figure S4) or for molecules with non-aromatic rings (rings with 4, 8, 12, ... atoms). Remember that this cancellation is sensitive to bond alternation, which pushes the interference feature away from $E = 0$.

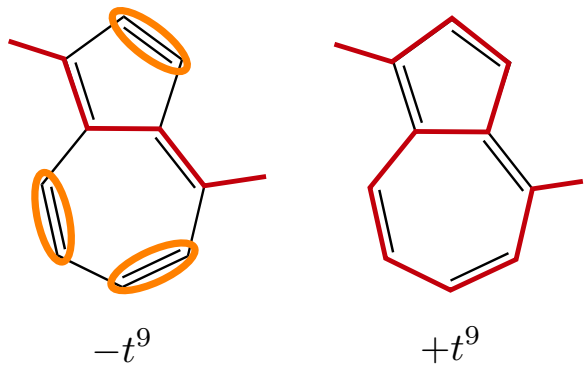

$-t^9$  $+t^9$

Figure S4: Azulene molecular junction showing the cancellation of the only two non-fix-point diagrams. This cancellation leads to destructive interference at $E = 0$.

The cross-conjugation rule of thumb still holds (for Hückel models) when considering

alternant hydrocarbons (possibly containing aromatic subsystems) as long as there are no closed loops outside any of the possible tunneling paths. However, when writing down the determinant polynomial, all the graphical rules must still be applied in order to obtain the correct number and the correct positions of the transmission nodes.


\end{document}